\RequirePackage{lineno} 
\documentclass[review,12pt]{article}
\usepackage{graphicx}
\usepackage{amsmath}
\usepackage{amsthm}
\usepackage{epstopdf}
\usepackage{setspace}
\usepackage[numbers]{natbib}
\usepackage[hyphens]{url}
\usepackage{hyperref}
\usepackage{float}

\theoremstyle{definition}

\theoremstyle{definition}
\newtheorem{remark}{Remark}[section]

\title{A Practitioner's Guide to Multiple Testing Error Rates}
\author{Jonathan Rosenblatt \\
	Department of Statistics and Operations Research,\\
	The Sackler Faculty of Exact Sciences, \\
	Tel Aviv University \\ 
	Israel}
\begin{document}
% \linenumbers
 \maketitle

\section{\label{sec:summary}Summary}
In this review, we present the problem of selecting the error rate, when considering multiple hypothesis testing. 
The statistical literature gives most attention to the error controlling procedures, but selecting the appropriate error rate is not a trivial task. 
We present the most common error rates, such FWER, FDR and their derivatives. We present the considerations involved in the choice of the error rate and demonstrate them using several examples from the fields of genetics, medical imaging, eduction policy and psychology.

\section{\label{sec:introduction}Introduction}

It is quite common in modern research for a researcher to test many hypotheses. 
The statistical (frequentist) hypothesis testing framework does not scale with the number of hypotheses in the sense that na\"{i}vely performing many hypothesis tests will probably yield many false findings;  ``false'' in the sense they will not be replicated.
Indeed, classical statistical ``significance'' is evidence for the presence of a signal within the noise expected in a single test, not in a multitude where the noise levels are higher. 
Strong evidence of signal assuming one noise level, can easily be considered as no evidence of signal under a higher noise level. 
For protection from an uncontrolled number of erroneous findings, a researcher has to consider the type of errors, or non-replications, he wishes to avoid. The researcher can then select the adequate procedure for that particular error type and data structure, or alternatively estimate that error type for a particular set of candidate findings. 

In practice, the selection of the proper error rate might cause the researcher some confusion. This point was made at the 2009 Multiple Comparisons conference in Tokyo \citep[][Section 4.4]{benjamini_simultaneous_2010}, demonstrated in the following question from the statistics Questions \& Answers web site \emph{Cross Validated}~\footnote{ See \url{http://stats.stackexchange.com/questions/26588/multiple-fdr-corrected-experiments-using-the-same-data}. Accessed on Apr 20, 2013}~:
\begin{quotation}\em
I am testing many (500,000) genetic variants, and the tests are FDR corrected and give me a q-value. Normally I would just call everything with $q < .05$ significant. But in this case I am testing those same genetic variants in two other related experiments (not using exactly the same individuals, but the samples may overlap). What to do? Would changing the significance threshold for $q$ to $.05/3=.0167$ be an option?
\end{quotation}
This particular example is further discussed in Section~\ref{sec:cross_validated}. 

To offer guidance, we review possible error types for multiple testing (sec~\ref{sec:measures_of_error}) and demonstrate them with some practical examples (sec \ref{sec:examples}) which clarify the formalism of sec~\ref{sec:measures_of_error}. Finally, in appendix~\ref{sec:on_your_pc}, we include some notes on the software implementations of the methods discussed.

A multiplicity control procedure (e.g. Bonferroni, Benjamini-Hochberg, \dots) is a data manipulation process-- an algorithm-- that guarantees that a preselected error rate is no larger than a preselected value. A typical procedures will actually offer guarantees vis-\`a-vis several error measures simultaneously. 
The emphasis of this manuscript is however on the error rates, and not on the multiplicity control procedures themselves.

For the purpose of selecting the appropriate procedure consult your favorite software's documentation (see our appendix~\ref{sec:on_your_pc}). Alternatively, \citet{farcomeni_review_2008} or more recently \citet{goeman_tutorial_2013}, can serve as  references. 
As the focus of this paper is the error measures, p-value adjustment, simultaneous confidence intervals, and error estimation will not be discussed. The reader is referred again to \cite{farcomeni_review_2008} or \cite{goeman_tutorial_2013} as possible references.

\section{\label{sec:measures_of_error}Measures of Error}

\subsection{Family Wise Error Rates}
Consider the testing of several null hypotheses against their respective research (alternative) hypotheses. The Family Wise Error Rate (FWER) is the frequency of experiments in which a false rejection of some null hypothesis will occur; put differently, the probability of a false finding.

As is customary in single hypothesis testing, a FWER level of $\alpha=0.05$ is often used and sometimes even required, as in drug registering experiments.

Table~\ref{tab:event_notation} introduces the nomenclature which has become standard in the multiple comparisons community and will be referenced throughout this article. Following this notation, the FWER is defined as 
$$Prob(V \geq 1 )$$ where $Prob(.)$ denotes the relative frequency over repeated experiments.

\begin{table}[h]
  \centering
\begin{tabular}{|c|c|c|c|}
\hline \rule[-1ex]{0pt}{1.5ex} & Claimed nonsignificant & Claimed Significant & Total \\ 
\hline
\hline \rule[-1ex]{0pt}{1.5ex} Null & $U$ & $V$ & $m_0$ \\ 
\hline \rule[-1ex]{0pt}{1.5ex} Nonnull & $T$ & $S$ & $m_1$ \\ 
\hline \rule[-1ex]{0pt}{1.5ex} Total & $m-R$ & $R$ & $m$ \\ 
\hline 
\end{tabular} 
  \caption{Classification of types of decisions made}
  \label{tab:event_notation}
\end{table}

\subsubsection{Weak and Strong Control the Family Wise Error Rate}
The probability of any particular inference procedure of making a false finding depends on the existence of true effects. FWER control in the ``weak'' sense refers to procedures which guarantee controlable FWER when there are no true effects at all, i.e., when all null hypotheses are correct. 

Detecting the existence of \emph{any} phenomena ($m_1>1$) is a simpler task that actually identifying these phenomena. Lack of weak FWER control means that we have no error guarantees even regarding this simple task. 
As mentioned in Section~\ref{sec:introduction}, a multiplicity control procedure might offer guarantees with respect to several error measure simultaneously. Weak FWER control should, and typically is, a minimal requirement. 

FWER control in the ``strong'' sense is the complementing concept, referring to procedures which guarantee FWER control even in the presence of some true effects, i.e., when not all null hypotheses are correct. 
As their names imply, ``strong'' control is the stricter criterion, which entails ``weak'' control.

\subsection{\label{sub:fdr}False Discovery Rates}

The False Discovery Rate (FDR), first introduced by \citet{benjamini_controlling_1995}, is the ratio between false discoveries and total discoveries, averaged over replicated experiments. 
Denoting the (unknown) False Discovery Proportion in a particular experiment as $ FDP=V/R $, and setting the convention that no discoveries signify no errors ($R=0 \Rightarrow FDP=0$), the FDR can be now defined as:
$$E \left( FDP \right)$$ 
where $E(.)$ denotes the average over all possible experimental results.

\begin{remark}
The FDR error rate has become synonymous with the Benjamini-Hochberg procedure presented in \citep{benjamini_controlling_1995} . This is plain wrong and confusing. Benjamini-Hochberg is a procedure that does indeed offer FDR control in particular setups, but it is only one of many.
\end{remark}

\begin{remark}
In the context of FDR, there need not be an $\alpha=0.05$ convention. Researchers are free to choose the level of error they see adequate for their particular research. Obviously, an error level of $\alpha=0.5$ might be hard to defend from critique. Having said that, since FDR control does offer FWER control in the weak sense, the $\alpha=0.05$ convention might be justifiable.
\end{remark}

\subsection{\label{sec:other_measures}Other Measures of Error}
FWER and FDR are the most commonly used, but by no means the only measures of error. Since many error measures are merely an average over replications of the experiment, many other error measures can be considered by replacing the error function to be averaged. Denoting by $E(C)$ the average, over all possible experimental outcomes, of some error $C$. We now see that FWER and FDR simply set  $C = I_{\{ V \geq 1 \} } $ and $C = FDP$ respectively.
Some other measures of error are:

\begin{itemize}

\item Per Family Error Rate (PFER): Where $C=V$.\\
This measure is the simple (expected) number of erroneous discoveries. 

\item Per Comparison Error Rate (PCER): Where $C=V/m$.

\item $k$-FWER \citep{van_der_laan_augmentation_2004}: Where $C(k) = I_{\{ V \geq k \} }$.\\
This measure is the relative frequency of the making of no less than $k$ erroneous discoveries.

\item False Discovery Exceedance (FDX)\citep{genovese_exceedance_2006}: Where $C(\gamma) = I_{\{ V/R \geq \gamma \} }$.\\
This measure was motivated by the fact that FDR keeps the proportion of false discoveries small, but only \emph{on average}. 
In extreme scenarios, FDR does not exclude the possibility of making more than $FDP>\alpha$ mistakes in \emph{almost all} experiments. FDX targets these scenarios explicitly, by allowing it to happen with a small probability, and is thus more conservative than FDR. 

\end{itemize}

Other measures of error which are not simple averages over replications of the experiment include, but are not limited to:
\begin{itemize}

\item Positive FDR or pFDR \citep{storey_direct_2002} or $FDR_{-1}$ \citep{benjamini_discovering_2010} : Defined as $E(V/R;R>0)$.\\
This measure is essentially the proportion of false findings (within all findings), but averaged, not on all possible experiments outcomes, but only on those which actually return findings. It was motivated by the observation that the FDR of a procedure might be very low merely because in many events it returns no findings, even if it makes many mistakes when it does indeed return findings (see \citep{storey_direct_2002} for a example). 
On the other hand, if all the null hypotheses are true, thus all findings are false, one would want an error measure to coincide with the probability of a false finding (weak FWER). pFDR does not enjoy this attribute.

\item Marginal FDR or mFDR \citep{sun_oracle_2007} or Fdr \citep{efron_microarrays_2008} or $FDR_{+1}$ \citep{benjamini_discovering_2010}: Defined as $E(V)/E(R)$. \\
While not very interesting for itself, this error measure gained popularity since it is mathematically tractable and approximates the FDR when many independent hypothesis are being tested. 

\end{itemize}

We conclude by noting that FWER, FDR, pFDR and mFDR are (currently) by far the most popular. So much so that it is actually hard to find published applied research using any other.

\subsection{Choosing Your Family}
All the previous error measures are defined for a family of hypotheses that is known, and indirectly assumed this family is all the hypotheses being tested in an experiment. This need not be the case, and defining the family of relevance may be a non-obvious part of the researcher's work. The examples in Section~\ref{sec:examples} include some trivial scenarios, in the sense that the family of hypotheses is clear. The section also includes some examples where the family is not trivial (see \ref{eg:imaging_genetics}) and its choice will depend on the scientific statement in mind.

\section{\label{sec:examples}Examples}

\subsection{\label{sec:tukey_exams}Tukey's Psychological Exams}
In his 1953 unpublished paper: ``The Problem of Multiple Comparisons'' \citep{benjamini_john_2002} and later, when lecturing at Princeton University \citep{donoho_higher_2004}, Prof. John Tukey would tell a motivating tale about a young psychologist. After administering 250 tests he finds that 11 were significant at the 0.05 level. A null hypothesis being that a given test does not differentiate between his groups of interest, a (significant) rejection of this null means a test does actually differentiate the groups.
After the initial feeling of satisfaction he consults a senior researcher, only to discover his findings are rather poor, since one  would expect 12.5 significant tests due to chance alone. Having only 11 significant results is actually disappointing.

With this new understanding, our psychologist now has to decide how he can protect himself from false findings. 
Say the tests consist of new candidate clinical diagnostics for condition X. Making an error means that a test will be used to diagnose X while it actually cannot distinguish between healthy and X. Since this is unacceptable for our psychologist, he will want an inference procedure that controls the FWER in the strong sense. This will also guarantee protection  in the case that no test differentiates between healthy and X, i.e., weak FWER control, which can actually be a question of interest for itself.

Now consider a different scenario: The tests check for differences in personality attributes between genders. Making an error means that the psychologist might believe male and female differ in a way they actually do not. The researcher does not consider this a serious mistake, as long as many other true differences are discovered. In this setup, the researcher should control the FDR, FDX or pFDR. 
Allowing for some mistakes will allow the researcher to enjoy a sensitivity gain compared to FWER-controlling-procedures. The interpretation of the findings should be done in accord with the error measure employed.

\subsection{\label{sec:anova}ANOVA}
In their \citeyear{williams_controlling_1999} paper, \citeauthor*{williams_controlling_1999}, analyze the National Assessment of Educational Progress (NAEP) 1990 and 1992 data. This data consists of the average eighth grade mathematics proficiency scores for the 34 states that participated in both 1990 and 1992 NAEP Trial State Assessment (TSA). Comparisons are made between regions (Central, Northeast, Southeast and West), years (1990,1992), states nested within regions, and the Year~$\times$~Region interaction. 
In this case a null hypothesis means there is no difference in the proficiency score between sub-groups, and its rejection meaning there is indeed such a difference.

We start by noting that this one study, provides four families of hypotheses. Indeed, a falsely discovered difference between regions, as an example, is of no concern when comparing years, states or regional changes.

We also note that in their paper, the authors actually compare between procedures controlling the FWER and the FDR. They do not offer a justification for the preference of any measure over the others, so we will offer one of our own. We will remark however, that their bottom line is unorthodox in the context of ANOVA: 
\begin{quote}\em 
Each of the three authors believes that the B-H procedure is the best available choice
%--- \citeauthor*{williams_controlling_1999}
\end{quote}

So why FWER? If, for example, the study is analyzed in the context of discrimination-- having no policy implications but rather a possible stigmatizing effect-- the researcher might wish to refrain from any falsely discovered differences between states, regions etc. If, on the other hand, intervention policies are the context, then power is a major concern. Missing a difference might mean policy makers are left unaware of the differences to be addressed. This context requires a less stringent error criterion than FWER, leading to the authors' stated preferences.

\subsection{\label{sub:fMRI}Functional Magnetic Resonance Imaging}

Consider now the case of the neuroscientist, trying to locate the brain regions responsive to visual stimuli. He has scanned a dozen subjects or so in the Magnetic Resonance Imaging (MRI) machine and recorded the brain's activation\footnote{ He actually measured the blood oxygenation level. Details can be found in \cite{lazar_statistical_2008}.} in response to the stimuli. To be precise, he measured the activation level at \emph{each} of several thousand brain locations, called Volumetric Picture Elements (voxels); their exact number depending on the resolution of the MRI scan . With the measured activation levels in hand, the researcher can compute their correlation to the stimulus given. If the voxel-wise measurement is correlated with the stimulus, the location is considered ``active''. We see that localizing activation actually consists of performing many local hypothesis tests: the null hypothesis of no correlation to the stimulus is tested at each voxel, and its rejection meaning a responsive location has been found.

Returning to multiplicity error rates; an error would mean the researcher declared a voxel as responsive when it actually is not. This does not seem like a terrible mistake to make, so the researcher should probably protect himself from large proportions of errors, and not from the making of one single error. FWER, k-FWER and Per Family Error Rate are thus excluded. Per Comparison Error Rate seems like a possible candidate, but it is very liberal. One can actually gain power by including many ``junk'' hypotheses, say, by including the air surrounding the head in the family. To see this, consider the case of infinitely many hypotheses tested. The proportion of errors will trivially be smaller than any $\alpha$ we pick. 

Our researcher is thus left with FDR and FDX as candidates for measurement of error. 
In the case the researcher has no clear favorite from within these two measures, a possible consideration at this stage might be the availability, simplicity and power of controlling procedures. These considerations give preference, at the time of writing, to FDR over FDX. 

\begin{remark}
For fairness it should be stated that ``cheating'' with FDR and FDR is possible, by augmenting the family with some obviously \emph{false} null hypotheses~\cite{finner_false_2001}. 
It is our view that ``cheating'' the FDR or FDX does require more effort and malicious intent than analyzing a needlessly large brain volume. We thus do not qualify these two ``cheats'' as equally problematic.
\end{remark}

\subsubsection{Functional Magnetic Resonance Imaging-- Cluster Level Inference}

Return to the neuroscientist from sec~\ref{sub:fMRI}. Recalling that the voxels are arbitrary volume units, defined by the technology of the MRI and not by entities of interest for inference, he decides that a more interesting entity is a mass of contiguous activations. He thus decides that he is interested in spatially contiguous regions with activation larger than ``7'' (in some scale). These regions are known as ``excursion regions'', ``exceedance sets'', ``blobs'' and possibly other names. 
After scanning a subject, he realizes there are 30 contiguous regions which exceed 7. Conscious that some are  due merely to chance variation, and knowing enough probability theory, he computes a p-value for the observed volume (exceeding 7), in each of the 30 regions. If he rather not make any mistakes, he can control the FWER of the regions. Namely, controlling the probability of declaring any inactive regions as active. This is indeed the approach implemented in several brain analysis software packages, particularly SPM (\url{http://www.fil.ion.ucl.ac.uk/spm/}).

Alternatively, if the researcher wishes to allow for some slack and accept false regions-- as long as their proportion is not too high-- he should use FDR or FDX control. Alas, the FDR defined in Section~\ref{sub:fdr} assumes an a priori fixed and known number of hypotheses being tested ($m$). The number of excursion regions is data dependent, thus random and a priori unknown. Extensions of the FDR for the random-number-of-hypothesis case do exist. A rigorous exposition can be found in  \citet{siegmund_false_2011}. Note however, that error-controlling \emph{procedures} for the random hypothesis case are not as abundant and studied as the fixed hypothesis case. The mathematical proofs would typically require some difficult to justify assumptions. Simulation performances however, do seem promising \citep{chumbley_false_2009,chumbley_topological_2010}.

\subsection{Functional Magnetic Resonance Imaging-- Clinical Scan}
Return again to the neuroscientist from sec~\ref{sub:fMRI}. This time his single patient is about to enter surgery for the removal of a brain tumor. 
The patient will be scanned in the fMRI in order to localize the speech regions, as the tumor is residing nearby and the surgeon needs to be extra-careful around these regions. 
In this clinical case there are different considerations than in basic scientific research. 
Type $II$ errors are arguably more important than type $I$ errors: underestimating the speech region might cost the patient his verbal skills; overestimating it, might cost him an extra surgery or a recurring tumor. 
None of the error measures presented until now is concerned with false negatives. 
Referring to the terminology in Table~\ref{tab:event_notation}, our neuroscientist would probably be interested in something like $E(T/(m-R))$, which captures the sensitivity of the inference. 
This measure is the False Non Detection Rate (FNR) \citep{genovese_operating_2002}. 
We have not presented this measure yet, as it is concerned with the \emph{non-detections}. We shall revisit it in the context of power in Section~\ref{sec:power}.

\subsection{\label{sec:GWAS}Genome Wide Association Studies}
In a typical Genome Wide Association Study (GWAS) the geneticist will record the genetic information of many subjects (genotyping) with the aim of discovering associations between the genotype and the individuals' attributes (phenotype). Assuming a univariate phenotype, the researcher will perform some type of regression between the phenotype and each genetic attribute (titled single nucleotide polymorphism-- SNP). With today's technology, the number of SNPs considered in a typical GWAS is hundreds of thousands. To declare an association, a researcher will try to reject the no association null hypothesis between \emph{each} SNP and the phenotype, leading to the simultaneous testing of several hundreds of thousands of hypotheses. 
Since the researcher does not concern himself with the making of a single mistake, as long as other associations discovered are true, he should choose FDR control or one of its relatives discussed in Section~\ref{sec:power}. 
That said, it is also very common in GWAS, to use a p-value threshold of $10^{-7}$. This threshold is intended for FWER control, when searching over $500,000$ SNPs and using the Bonferonni procedure \cite{bush_chapter_2012} for FWER control. 

So FDR or FWER? It is left for the researcher to decide, and it ultimately depends on the implications of declaring false associations.

\subsection{\label{sec:cross_validated}Cross Validated Example}

In this example, the researcher is looking for associations between SNPs and three distinct\footnote{ It is actually implicit whether these are distinct phenotypes or not. We have assumed they are distinct, because of the ``subject overlap'' comment.} phenotypes. The error measure has already been selected. The family groupings are unclear. 
The options being (a) accounting for errors only within each experiment, leading to three families of hypotheses, or (b) global error accounting, leading to a single family. 

Both approaches have their advantages and disadvantages. 
By keeping the experiments separate we gain power but the global error rate is no longer $\alpha$. \citet{yekutieli_hierarchical_2008} has approximated, for some cases, that under strategy (a) with $\alpha$ level FDR control within each experiment, then the \emph{global} FDR should actually be close to 
\begin{equation} 
	FDR \approx \alpha \cdot \frac{\text{Total discoveries} + 
	\text{No. of experiments}}{\text{Total discoveries}+1} \label{eq:families_FDR}
\end{equation}

Eq.~\ref{eq:families_FDR} captures the intuition that the more experiments performed while controlling only for errors within the experiment, the global error rate might inflate. 

The discussed problem includes three experiments, assumingly looking for three different phenomena. The fact we discuss \emph{three} experiments, which were all conducted by the same researcher, is quite arbitrary. Why not control for the errors performed in the whole of science? Or at least in all genetic association studies. 
A proper discussion of this matter requires an unplanned detour into the philosophy and sociology of science, and is not part of this guide. 
We will conclude by remarking that combining errors over different phenomena is indeed  desirable~\cite{ioannidis_why_2005}, yet rarely performed in practice.

\subsection{\label{eg:imaging_genetics}Imaging Genetics}

The field of imaging genetics aims at finding the genetic attributes associates with phenotypes derived from medical imaging. In a pioneering study, \citet{stein_voxelwise_2010} set out to find the genetic variation associated with local brain volume, under the paradigm that different genes affect different brain regions. 
The data included the genotyping and imaging of $N \approx 700$ individuals. 
The genotype of each individual comprises information of $n_G \approx 400K$ SNPs. 
The imaging data encodes the relative volume of each subject at $n_B \approx 30K$ voxels. 
Testing for association between all $\{SNP\} \times \{voxel\}$ combinations, leads to $n_G \cdot n_B \approx 12\times10^9$ hypotheses. Should they all be considered one family of hypotheses? Or maybe each SNP (or voxel) is actually a separate family? 

A researcher might want to infer which gene is associated with which location.
A single family of hypotheses will include all $\{SNP\} \times \{voxel\}$ combinations. 
FWER control over this family is out of the question. 
FDR control means that the researcher is concerned with the proportion of false associations detected within all of the $\{SNP\} \times \{voxel\}$ associations found. This seems like a good criterion, except for the fact it requires correcting for $12\times10^9$ hypotheses. 
Our researcher starts considering alternative error criteria. 
A natural option might be SNP-wise testing, perhaps using the B-H procedure over all voxels within each SNP. Power is certainly gained as each family is corrected only for the $n_B$ voxels within it. 
What about false findings? Sadly, this approach offers no error control. To see this, consider the case where there is only one voxel: this amounts to $n_G$ level $\alpha$ hypothesis tests, which is this initial multiplicity problem. 

A more justifiable solution might harness the hierarchy of the problem; that is, by selecting associated SNPs and localizing the association only for these selected SNPs. 
Naturally, the multiplicity is alleviated since only selected SNPs will be passed for voxel-wise testing. However, if the same data is used for selecting the SNPs and then selecting the voxels, an $\alpha$ level FDR control within selected SNPs will still not guarantee an $\alpha$ level FDR control, across discovered $\{SNP\} \times \{voxel\}$ associations. This is actually a case of \emph{selective inference} \citep{benjamini_simultaneous_2010}, also referred to by practitioners as ``data snooping'' or ``double dipping''. 

Having given it more thought, our researcher decides she wants a method that has two properties: 
(a) controlling for the number of falsely discovered SNPs; and 
(b) controlling for the number of falsely discovered voxels associated with each discovered SNP. 

To put if formally, Table~\ref{tab:event_notation} needs refinement. 
Define $R$ and $V$ to be the number of discovered SNPs and \emph{falsely} discovered SNPS respectively. 
Define $R_g$ to be number of voxels declared associated with SNP $g$, and $V_g$ accordingly. The desired measure of error has two requirements: 

\begin{equation} \label{eq:hirarchial_error}
 E \left(\frac{V}{R} \right)\leq \alpha_1 
\text{ and } 
E \left( \frac{1}{R}\sum_{g} \frac{V_{g}}{R_{g}} \right)\leq \alpha_2
\end{equation}

Is there a procedure that controls this type of error? While novel and under active research, there is presently one such procedure\footnote{ With proofs for the independent test-statistics case}. It has two stages. 
First, the omnibus-stage:  testing for an associated SNP by aggregating over voxels within SNP and controlling for the number of SNPs tested. Second, a post-hoc stage: drilling into the selected SNPs searching for associated voxels. The novelty of the procedure is at the second stage, which controls for the number of voxels with a conservative error rate, which accounts for the previous SNP selection stage. 
The details can be found in \cite{benjamini_adjusting_2013}.

\section{Simultaneous versus Selective Inference}
Up to this point, we have motivated the choice of error measure by a mere ``error accounting''. 
There is actually another perspective, which can make the choice of the error measure quite obvious once it has been recognized. 
As put by \citet{cox_remark_1965}:
\begin{quote}\em
It might be better to talk about the problem of selected comparisons rather than about the problem of multiple comparisons.
\end{quote}

Cox's insight was that making statements on the truthfulness of a subset of selected hypotheses, and making statements about the simultaneous correctness of a subset of hypotheses is not the same thing. 
Think in terms of replicability: replicating a combination of phenomena is not the same as replicating each separately.
Naturally, the simultaneous truthfulness of the selected hypotheses, entails the truthfulness of each and every one of them. Thus, simultaneous inference is the more ambitious task.
In  Cox's words \cite{cox_remark_1965}:
\begin{quote}\em
The fact that a probability can be calculated for the simultaneous correctness of a large number of statements does not usually make that probability relevant for the measurement of
the uncertainty of one of the statements. If we are directly interested in a single statement about the vector parameter, the probability of simultaneous correctness would, however, be appropriate. The practical usefulness of the multiple comparison techniques then usually lies in giving a conservative bound for the effect of selection, rather than in giving an ``exact'' solution.
\end{quote}

Armed with the distinction between simultaneous and selective inference, we can relate error measures to inference types. 

Simultaneity ambitions are only satisfied with FWER control. This is simply because allowing any errors to filtrate, ruins the truthfulness of the combination of claims.
FDR is clearly a selective, and not simultaneous statement. It can be seen as the average truthfulness of the selected hypotheses. Put differently, FDR is actually controlling the PCER  within the selected hypothesis subset.
FDX is also selective. It ensures a high proportion of truthful statements within the selected hypotheses. 

Demonstrating using the examples: 
The background in Tukey's psychological exams from Section~\ref{sec:tukey_exams} was too vague to determine which inference type is appropriate. Both can be advocated. 
The same goes for the NAEP example from Section~\ref{sec:anova}. 
Assuming the neuroscientist from the fMRI example in Section~\ref{sub:fMRI} cares of the truthfulness of each detected location, and not their particular combination, this is a case of selective inference. A similar consideration holds for the case of associated SNPs in the GWAS example in Section~\ref{sec:GWAS} and $\{SNP\} \times \{voxel\}$ associations in Section~\ref{eg:imaging_genetics}.

\begin{remark}
Although not the in scope of this manuscript, it is only appropriate to mention False Coverage Rate adjusted confidence intervals. These interval estimators, suggested by \citet{benjamini_false_2005}, make the distinction between selective and simultaneous clear. They are not simultaneous, as they do not offer joint coverage of the selected parameters. They do however offer a predefined average coverage over the \emph{selected} parameters.
\end{remark}

\section{\label{sec:power}Power Considerations}

The reader might have noticed that the different error measures in Section~\ref{sec:measures_of_error} care only of the number of discoveries and false discoveries. Our interest in detection sensitivity is naturally implicit in the procedures researchers employ. Otherwise, never rejecting any null hypothesis, will trivially control all of the error types in Section~\ref{sec:measures_of_error}. 
Power can benefit from (a) knowledge of the proportion of signals in the noise ($ 1 - m_0 / m $) or (b) from an introduction the expected deviations from the null hypotheses. 

To demonstrate (a), consider two researchers doing the same research. The first, which did some more reading on the topic, knows with complete certainty that there are 10 false null hypotheses (signals) in his 100 hypotheses tests. The second, being less through, has no access to this knowledge. 
Naturally, the first can exploit this information. As a trivial example, he will know that  more than $10$ rejections will certainly contain errors\footnote{ This does not mean that the first $10$ are necessarily true. Only that more than $10$ will certainly contain errors.}. 

To demonstrate (b), consider a scenario where the researcher is certain that if an effect exists, it would be of magnitude $\pm 7$ (in some arbitrary scale, say z-values). Performing a one-sample $z$ or $t$ test, while ignoring this belief, will lead the researcher to reject all hypotheses with large (absolute) effects. Particularly, an effect of, say 20,  will be considered very extreme, with infinitesimally small p-values. But, when considering the fact that effects are expected to be near 7, the researcher might actually prefer to reject effects near 7 \emph{before} he rejects effects near 20. In  statistical terminology, this is simply an underpowered test constructed for the wrong alternative hypothesis. 

Specifying the expected deviation from the null for each hypothesis tested is no easy task. There exist however, several procedures which use the multitude of hypotheses tested in an attempt to empirically characterize the deviations from the null~\footnote{ Under mild assumptions regarding the form these deviations might take. Essentially assuming that deviations from the null are not uniformly dispersed but rather tend to clump together.}, and harness this information to gain power. 
Essentially all rely on estimators of the probability of a hypothesis being a true null given the value of some test statistic $z_i$. This is the posterior probability of the null, also named ``local fdr'' and denoted by $fdr(z_i)$. Details can be found in \cite{efron_microarrays_2008}. Also see Remark~\ref{sec:baeysian_interpretation} on the existence and interpretation of this probability.

This magnitude-- the probability of being null given the data-- is not an error rate but rather a test statistic. It is however a rather intuitive test statistic. So much so that many authors set the rejection criterion to, say, $fdr(z_i)\leq 0.2$. This lends itself to the interpretation that results with frequencies smaller than 1 in 5, under the null assumption, are ``dangerously prone to wasting investigators' resources''~\cite{efron_microarrays_2008}.

\citet{storey_positive_2003} establishes a relation between $fdr(z_i)$ and the Marginal FDR from Section~\ref{sec:other_measures} so that a researcher opting for the $fdr(z_i)\leq 0.2$ criterion, can receive some sense of how many errors per discovery he will be doing on average.  The relation is data dependent. In the problem analyzed by \citet{efron_microarrays_2008}, rejecting for $fdr(z_i)\leq 0.2$, is approximately equivalent to marginal FDR control of $0.1$.
This relation is even more appealing, since with a growing number of hypotheses being tested, the marginal FDR is a good approximation of the FDR. 

Returning to power considerations, and using the notation presented in Table~\ref{tab:event_notation}, we can specify many error measures which capture the idea of maximal power subject to the false detections being kept at a low level:

\begin{align}
        min\{FNR \quad \text{such that} \quad FDR\leq \alpha \} \label{eq:compound_1}\\
	min\{mFNR \quad \text{such that} \quad mFDR\leq \alpha \} \label{eq:compound_2}\\
	min\{T \quad \text{such that} \quad V \leq \alpha \} \label{eq:compound_3}
\end{align}

The different procedures aimed at satisfying these error functions typically use the local fdr statistic ($fdr(z_i)$) as a test statistic, but differ in the heuristics used to  compute this statistic~\cite[eg.][]{storey_direct_2002,efron_microarrays_2008,sun_oracle_2007}, typically, assuming a large number of hypotheses being tested and independence between test statistics. 

The search for procedures with desirable properties under errors measures such as eq~\ref{eq:compound_1}  - eq~\ref{eq:compound_3}, is ongoing. It is an active field of investigation with beautiful theory, developed either by the Multiple Comparisons Community or borrowing from statistical decision theory \cite[see][]{sun_oracle_2007}. The analysis of the finite-sampling properties of suggested procedures with respect to some desirable error measures is not an easy task.  For a complete treatment, and state of the art procedures, the reader is referred to \cite{efron_large-scale_2010}.

\begin{remark}\label{sec:baeysian_interpretation}
Assigning a \emph{posterior} probability to a hypothesis being null requires also assigning a \emph{prior} probability to this event. The interpretation of this probability has raised some controversy \cite[e.g.][]{morris_comment:_2008} as it can be seen as a statement of subjective beliefs, of sampling frequencies or merely as descriptive. Settling this interpretation issue is outside the scope of this manuscript.
\end{remark}

\section{Acknowledgments}
I wish to thank the many friends and colleagues who contributed of their time and knowledge, to improve clarify this manuscript: 
Prof. Yoav Benjamini, Dr. Daniel Yekutieli, Dr. Jelle Goeman, Dr. Aldo Solari, Kornelius Rohmeyer, David Golan, Neomi Singer and Samuel Cohen.

\bibliographystyle{plainnat}

\bibliography{Project-PracticalGuideToMultiplicity}

\appendix

\section{\label{sec:on_your_pc} On Your Computer}
It does not suffice to choose an error measure in order to perform an analysis. An error-controlling procedure will also have to be chosen, and this is what you should look for in your favorite software. This might be a general purpose statistical suite, or a problem-specific application. In the latter case, we have little to suggest, as domain specific applications typically implement the procedures popularized in that field. 
In the brain imaging example, popular software include SPM, Brain Voyager, FSL, AFNII. All incorporate the multiplicity control procedure preferred by their authors (thus implicitly, the error measure). 
In the GWAS example, the same occurs in software such as Plink, PRESTO, PERMORY and others. 
General purpose statistical software are built for flexibility in the analysis, and thus incorporate more multiplicity control procedures. 

In the R programming environment \citep{r_development_core_team_r:_2011} the function \emph{p.adjust} in the \emph{stats} package will allow you to perform the most common procedures. The references in the function documentation are a good starting point for learning about these procedures. 
For FWER controlling procedures, in particular in the context of linear contrasts in regression models, the \emph{multcomp} package is a good option. 
For FDR control (and variants) many packages have been written. A good listing of these can be found in \citet{bretz_multiple_2010} or Korbinian Strimmer's web site: \url{http://strimmerlab.org/notes/fdr.html}. 
We also note that, to the best of our knowledge, the hierarchical testing scheme in Section~\ref{eg:imaging_genetics} has not been implemented. Its implementation would require a new syntax to describe the hypotheses' hierarchy (families) and it is an open challenge. 

In SAS, multiple testing procedures are incorporated within PROC MIXED and also in PROC MULTEST.  The canonical reference is \citet{westfall_multiple_2011}. 

In SPSS, multiplicity corrections are typically found as part of the \emph{post-hoc} options of the analysis methods.

\section{\label{sec:glossary} Glossary}
Some of the terms in the multiple-comparisons literature, have appeared in several other disciplines under different names. To ease the transition, and for completeness, we present a glossary. 
In this glossary, we use as a reference the statistical nomenclature in Table~\ref{tab:event_notation}. Also note that we use \emph{rate} for the average of a \emph{ratio} or a \emph{proportion}. The literature is not consistent regarding this convention, so that the terms might be found in use for both purposes.

\begin{table}[H]
\centering
\caption{Glossary}
\begin{tabular}{{llp{9cm}}}
\hline \rule[-1ex]{0pt}{1.5ex} & Symbol & Names  \\ 
\hline
\hline \rule[-1ex]{0pt}{1.5ex} & $S$ & True Positives, Hits, True Discoveries  \\ 
\hline \rule[-1ex]{0pt}{1.5ex} & $U$ & True Negatives  \\ 
\hline \rule[-1ex]{0pt}{1.5ex} & $V$ & False Positives, Type $I$ Errors, False Discoveries, False Alarms  \\ 
\hline \rule[-1ex]{0pt}{1.5ex} & $T$ & False Negatives, Type $II$ Errors, False Non Discoveries, Misses  \\ 
\hline \rule[-1ex]{0pt}{1.5ex} & $V/R$ & False Discovery Proportion (FDP), False Discovery Ratio, False Detection Ratio/Proportion, False Alarm Ratio/Proportion, False Positive Ratio/Proportion, Fall-Out  \\ 
\hline \rule[-1ex]{0pt}{1.5ex} & $E(V/R)$ & False Discovery Rate (FDR), False Detection Rate, False Alarm Rate, False Positives Rate  \\ 
\hline \rule[-1ex]{0pt}{1.5ex} & $R/m$ & Accuracy  \\ 
\hline \rule[-1ex]{0pt}{1.5ex} & $E(T/(m-R))$ & False Non Discovery Rate (FNR)  \\ 
\hline \rule[-1ex]{0pt}{1.5ex} & $T/(m-R)$ & False Non Discovery Ratio  \\ 
\hline \rule[-1ex]{0pt}{1.5ex} & $S/R$ & Positive Predictive Value (PPR), Precision, Hit Ratio  \\ 
\hline \rule[-1ex]{0pt}{1.5ex} & $E(S/R)$ & Hit Rate  \\ 
\hline \rule[-1ex]{0pt}{1.5ex} & $U/(m-R)$ & Negative Predictive Value  \\ 
\hline \rule[-1ex]{0pt}{1.5ex} & $E(T)/m_1$ & Non Discovery Rate  \\ 
\hline \rule[-1ex]{0pt}{1.5ex} & $T/m_1$ & Non Discovery Ratio  \\ 
\hline \rule[-1ex]{0pt}{1.5ex} & $S/m_1$ & True Positive Ratio, Recall, Average Power  \\ 
\hline \rule[-1ex]{0pt}{1.5ex} & $U/m_1$ & Specificity, True Negative Ratio  \\ 
\hline \rule[-1ex]{0pt}{1.5ex} & $E(U)/m_1$ & True Negative Rate  \\ 
\hline \rule[-1ex]{0pt}{1.5ex} & $E(S/R)$ & True Positive Rate  \\ 
\hline 
\end{tabular}  
\end{table}

\end{document}